\documentclass[11pt,a4,onecolumn]{article}

\usepackage[english]{babel}

\usepackage{epsfig}
\usepackage{dcolumn}
\usepackage{amsmath}
\usepackage{changebar}
\usepackage{graphicx}

\newcommand{\ba} {\begin{eqnarray}}
\newcommand{\ea} {\end{eqnarray}}
\newcommand{\sun}{\ensuremath{\odot}}
\usepackage{graphicx}% Include figure files
\usepackage{dcolumn} % Align table columns on decimal point
\usepackage{bm}      % bold math
\usepackage{umlaut}
\begin{document}
\renewcommand{\figurename}{Fig.}
\renewcommand{\tablename}{Tab.}
\title{
  \vspace*{-4cm}{\normalsize\bf\hfill ECTP-2008-3} \\  
  \vspace*{2cm}
Cosmological Consequences of QCD Phase Transition(s) in Early
  Universe\thanks{Invited talk given at DSU08, ''Dark Side of the Universe'', Cairo-Egypt, June 1-5, 2008} }

\author{A.~Tawfik\thanks{atawfik@mti.edu.eg} \\
 {\small ECTP, the Egyptian Center for Theoretical Physics\thanks{http://ectp.mti.edu.eg/}}\\ {\small MTI Modern University, Mukattam, Cairo - Egypt }  
}

\date{}
\maketitle

\begin{abstract}
We discuss the cosmological consequences of QCD phase transition(s) on the
early universe. We argue that our recent knowledge about the transport
properties of quark-gluon plasma (QGP) should thraw additional lights on the
actual time evolution of our universe. Understanding the nature of QCD phase
transition(s), which can be studied in lattice gauge theory and verified in
heavy ion experiments, provides an explanation for cosmological
phenomenon stem from early universe.        
\end{abstract}

%\noindent
%PACS: , \\
%\hspace*{14mm}

\section{\label{sec1}introduction}
Our knowledge about the {\it early} and {\it present} universe is based on two
successful models; the standard model for elementary particles and the one
for cosmology. The main ideas about the early universe cosmology seem to be
confirmed from various cosmological data, like the temperature anisotropy in
cosmic microwave background radiation (CMB), the light element abundances,
etc. Although we have a solid description for phase transitions in almost all
epochs in the early universe, their confirmations still indirect processes. The
observation of cosmological 'exotic' phenomenon and the fruitful
results from heavy ion experiments are promising tools to confirm
these theoretical predictions.   

In this work, we limit our the discussion to QCD epoch, i.e. to energy scale
of $\Delta_{QCD}\approx 200~$MeV, $t\approx 10^{-5}-10^{-6}~$second after the
begin and when the universe has the {\it Hubble} radius of about $d_H\approx
10~$Km corresponding to scales of about $1~$pc~\footnote[1]{1 parsec
  $=3.08568025 \times 10^{16}~$m, the distance from the Sun which would result
  in a parallax of 1 arcsecond as seen from the Earth.} or about $3~$ light
years today. At high energy the coupling between quarks and gluons likely
becomes weaker. This coupling diverges at $\Delta_{QCD}=200~$MeV. From this
property, QCD is known as asymptotically free theory.   

As the energy decreases, i.e. the universe expands and cools down, the quarks
and gluons underwent to confined state, to hadrons (deconfinement-confinement
phase transition) and the chiral symmetry almost simultaneously breaks down,
i.e., orientation of right- and left-handed quarks. The latter likely leads to
production of Goldstone bosons. The three pions are the lightest Goldstone
bosons. Dynamical fluctuations on these particles and/or disoriented chiral
condensates (DCC) would manifest this transition.

In this talk, we discuss the cosmological consequences of nature and order of
the QCD phase transition(s) on early universe.

\section{Ideal and non-ideal Quark Gluon Plasma}

A theoretical framework for dynamics of phase transition(s) from quark-gluon
plasma (QGP) to hadrons still fails. In heavy ion collisions, we can 
detect produced particles at final state, i.e. after chemical and thermal
freeze-out. Consequently, there is no {\it smoking-gun} signal for occurrence
of equilibration processes while the energy density $\epsilon$ decreases. Same
statement is also valid for the degrees of freedom (d.o.f.). It is known
that the thermal (statistical) models, like hadron resonance gas,  work very
well in final 
state~\cite{TawRt,TawPT1}. These models presume a charge-conserved hadronic
phase, but they have no access to the phase transition itself.      

Although the energy density $\epsilon$ available to heavy ion experiments at
CERN-SPS~\cite{spsE}, BNL-RHIC~\cite{rhicE} and recently to
CERN-LHC~\cite{lhc1} exceeds the critical value calculated in lattice QCD
($\epsilon_c\approx 2\;$GeV$/$fm$^{-3}$ for physical quark masses and
vanishing net baryon number density), there is no unambiguous
QGP-signature. For example, $\epsilon$ achieved at RHIC $\sim5.5~$GeV/fm$^3$
for proper time $\tau_0\sim1~$fm/c. If 
$T_c\approx T_{fo}$, where $T_c$ is the critical temperature and $T_{fo}$ is
the freeze out temperature, holds at small or vanishing net baryon number
density, as we used to assume for ideal QGP, the phenomenological signals
characterizing the phase transition(s) have to remain measurable at final
state, at least the ones which are not sensitive to strong interaction or to
the medium, such as photons, leptons or color screening as $J/\Psi$ dissociation
into two leptons. According to recent lattice QCD results with almost
physical quark masses, $T_c\approx 200~$MeV~\cite{latticenew}. This value
exceeds $T_{fo}$ with at least $\sim 30~$MeV. During the relaxation time,
within which the system (universe) loses $\sim 30~$MeV most of dynamical
fluctuations might be moderated. According to~\cite{foTc23}, the inelastic
scattering rate should have steep $T$-dependence.

The plasma, like QGP, is a state in which the charges are screened due to the
existence of other mobile charges. This will modify the Coulomb's law. Lattice
QCD results in that the $J/\Psi$ bound state would survive up to $2T_c$, where
$T_c\approx270~$MeV~\cite{tawbb}. The RHIC results suggested to include
viscous corrections to the hydrodynamic evolution~\cite{rhicvisc1}, so that
the plasma state turns to be a fluid rather than an ideal gas of massless
components (quarks and gluons) and negligible correlations (free
interactions). 

Taking into consideration shear viscosity in QGP fluid leads to slower
hydrodynamic evolution relative to QGP with zero viscosity. The transverse
expansion in viscous QGP fluid turns to be stronger than the one in an ideal
QGP fluid. Also the particle production might considerably be enhanced in
viscous QGP fluid. That QGP turns to be viscous fluid, apparently, modifies
our ideas even about the time evolution of early universe.     

\section{QCD Phase Transition(s) and Chemical Freeze out}
\label{sec:qcd}

The QCD predicts that the asymptotically free quarks and gluons are weakly
correlated. QCD has been extensively studied on lattice for the last thirty
years; QCD Lagrangian has to be discretized. Then we put everything on a
finite a space-time lattice. It has been shown that a rapid change in various
thermodynamic quantities undoubtedly takes place at sufficiently high
energies~\cite{TawLat}. The degrees of freedom markedly decreases in a
relative narrow region of temperatures.  

First order phase transition at $T_c\sim 270~$MeV is evidenced in lattice QCD
without dynamical quarks (quenched)~\cite{Engels,fugu}. Including the dynamical
quarks makes the order of phase transition and the value of $T_c$ depending on
the mass and number of quark flavors.  
 
For two massless light quarks ($m_u=m_d=0$) and infinity heavy strange quark
($m_s\rightarrow\infty$) at vanishing net baryon number (chemical potential
$\mu_q$), which is appropriate to the early universe, where
$\mu_q/T\sim10^{-8}$, the phase transition is of second order and
$T_c\sim175~$MeV. If the light quarks get small masses, the critical behavior
of phase transition moves to smooth cross over. For degenerate quarks
($m_u=m_d=m_s=0$), the phase transition is again first order but with $T_c\sim
155~$MeV. For light up and down quarks and massive strange quark, the
transition is cross over and $T_c\sim 170~$MeV again. Therefore, we would
expect an upper value for the quark mass to secure first order phase
transition. Above this values the order of phase transition seems to be cross
over or weak second order.

It has been observed that the chiral symmetry is
restored at the same critical temperature $T_c\approx 154-174\;$MeV (depending
on the quark flavors) as that of the deconfinement phase transition. The
restoration of chiral symmetry means that the effective mass of quarks forming
the confined hadronic states becomes zero. Another important consequence of
chiral symmetry breaking restoration is the disappearance of mass degeneracy of
hadronic states having same spin but different parity quantum numbers. 

We find that the bulk thermodynamic quantities at very hight temperatures of
$4-5T_c$ remains below the Boltzmann limit~\cite{Peikert00,karsch07};
$\epsilon_{SB}\approx g \pi^2 T^4/30$. This behavior would indicate that
thermodynamic quantities may remain constant at much higher temperatures. This
means that the deconfined matter might remain {\it strongly} correlated. That
QGP at $4-5T_c$ turns to be strongly correlated, would have consequences on
its hydrodynamic evolution, production of confined hadrons and might delay the
freeze out processes.

The chemical freeze out is characterized by $s/T^3=7$ for three quark
flavors~\cite{TawRt}. $s$ is the entropy density. This condition characterizes
a stage at which annihilation and production processes are in chemical
equilibrium. As shown earlier, the consequence that $T_{fo}$ is below $T_c$,
would lead to steep $T$-dependence of the inelastic scattering rate.

\section{Cosmological Consequences}
\subsection{Nature and Order of QCD Phase Transition(s)}
At temperatures higher than the QCD critical temperature, $T_c$, the matter
are mainly formed in quarks, gluons, leptons and photons. In QCD epoch, the
energy scale ($\Lambda_{QCD}\sim 200~$MeV) is much larger than the
physical masses of these components, we therefore can approximate their masses
to be almost vanishing. As given above, the universe in this epoch has a
radius of $\sim10~$km, which leads to mass content of $\sim 4\pi
R_H^3\epsilon(T_c)/3 \sim 1.25M_{\sun}$. $\epsilon(T_c)$ is the energy density
at the critical temperature $T_c$. Its value is taken from lattice QCD
simulations~\cite{karsch07}. Under these circumstances the matter can be
treated as a radiation. 

The relaxation time scale of particle interaction at QCD energy scale
($\Gamma_q\sim\alpha_s^2T$ and $\Gamma_g\sim-\alpha_sT\ln g$, where
$\Gamma=t^{-1}$. $g$ and $\alpha_a$ being gauge and strong coupling constant,
respectively) is much shorter than the Hubble radius $H$, thus the different
phases of the matter; QGP, hadron gas, leptons and photons, are likely in
thermal and chemical equilibrium. Therefore this matter is much similar to
radiation fluid and the effective d.o.f are the baryon quantum numbers. Under
these conditions, the baryon number density can be calculated from   
\ba
n_B(T_c) &=& \eta~\cdot~s(T_c)~\left.\frac{n_{\gamma}}{s}\right|_{BBN} 
\ea
where $n_{\gamma}$ is the photon density at Big
Bang Nucleosynthesis (BBN). $\eta$ is the celebrated ratio of baryon density
asymmetry ($n_B-n_{\bar{B}}$) to photon density. According to recent WAMP
data~\cite{wamp1}, $\eta$ reads   
\ba
4\times 10^{-10}< & \eta & <7\times 10^{-10}
\ea
Using recent lattice QCD results~\cite{latticenew} for $T_c$ and $s$, we
can calculate the physical units of baryon number density $n_B(T_c)$ for three
dynamical quark flavors at physical masses in QCD epoch  
\ba
n_B(T_c) &=& 8.1\times10^{-11}~s(T_c)= 5.5\times 10^{-3}~MeV^3 
\ea
To get this value, we assume that no annihilation process is effective. The
number of baryons filling out the Hubble volume up to the causal horizon at the
QCD phase transition reads 
\ba
\left.N_B(T_c)\right|_H &\approx&  3\times 10^{48} 
\ea
Obviously, this number strongly depends on nature and order of QCD phase
transition as $s$ and $T_c$ do (review Sec.~\ref{sec:qcd}). When the universe
loses $\sim30~$MeV temperature, i.e. at the chemical equilibrium freeze
out~\cite{TawRt}, $N_B(T_c)$ consequently decreases to one half or one third of
this value.  

In first order phase transition, the two phase co-exist and bubbles ''dirt
objects'' in form of mixed phase of QGP and hadron gas are quite likely
expected. To keep the temperature constant, bubbles slowly dominate the
system. Since the two phases have different entropy densities, the bubbles
release latent heat and accordingly the energy density $\epsilon$
decreases~\cite{TawLat} until the transition is completed. Released latent
heat can reheat the matter (universe) to tiny fraction. Candidates for such
''cosmic'' bubbles would be primordial monopoles, cosmic strings, black holes,
etc. As there is so far no verification for any of these objects, we have to
look for other consequences, like primordial temperature
fluctuations~\cite{schwarz} and inhomogeneities forming dark matter clumps,
etc.  

At QCD phase transition, neutrinos move freely with mean free path
$\lambda_{free}\approx~10^{-6}R_H\approx~1~$cm and the fluctuations on their
diffusion scale might be washed out during QCD phase transition,
$\lambda_{diff}\approx~10^{-4}R_H\approx~1~$m. These scales determine whether
the temperature fluctuations (inhomogeneities) have to be neglected and the
bubble hadronization proceeds in a homogeneous rate or the bubble
hadronization is indeed inhomogeneous. According to~\cite{schwarz}, the second
scenario can be fulfilled, if the rms temperature fluctuations $>10^{-5}$. In
that case, the scale of inhomogeneity in baryon distribution~\cite{witten} can
be related to the scale of inhomogeneity in radiation fluid after the QCD
phase transition. This might modify the neutron to proton ratio and thus
explain the initial conditions for BBN. BBN is an important tool to
verify our ideas about the early universe. Inhomogeneities would lead to
formation of cold dark matter clumps in connection with the Hubble scale as
given above. The scale separation between bubble hadronization and hadron
diffusion is about two orders of magnitude. This scale difference has been
calculated for first order phase transition. It depends on released latent
heat and free energy difference between the two phases (QGP and hadron gas).   

\begin{figure}[thb]
\centerline{
\includegraphics[width=12.cm]{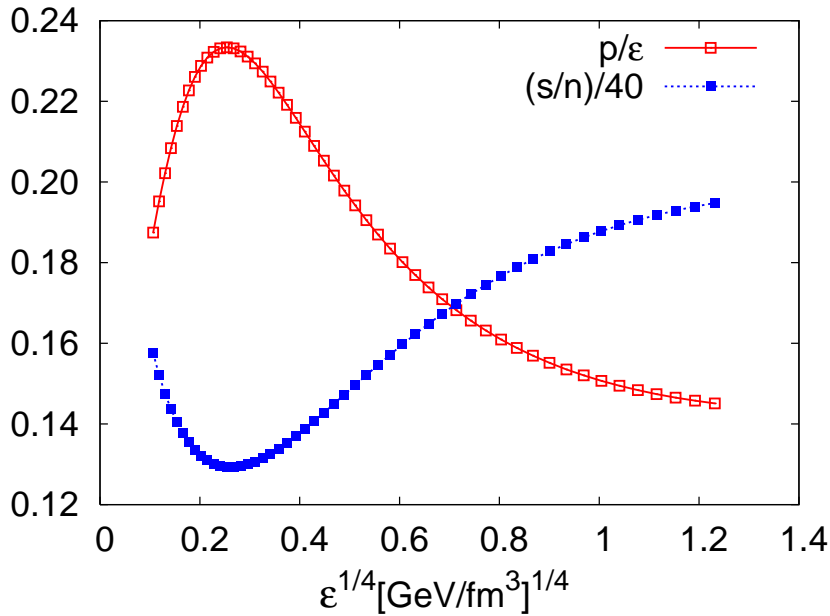}
}
\caption{It shows the results from a hadron resonance
  gas~\cite{TawLat} below $T_c$. Open symbols give $p/\epsilon$ as a function
  of $\epsilon^{1/4}$ in physical units [GeV/fm$^3$]$^{1/4}$. Solid symbols
  show normalized $s/n$. }   
 \label{Fig:cs}  
\end{figure}

Once again, in the first order phase transition, both phases exist at one
critical pressure and therefore the speed of sound $c_s^2=\partial p/\partial
\epsilon$ vanishes throughout the whole transition. During the time of
transition, the density perturbation within visible horizon ($\lambda<R_H$)
falls freely, i.e. radiation fluid velocity remains constant. At $T>T_c$, the
density perturbation modes causes acoustic oscillations. They entirely vanish,
when $c_s$ vanishes. At very high energies, $c_s=1/\sqrt{3}$ in
radiation-dominant matter (universe)~\cite{cs1}. Fig.~\ref{Fig:cs} shows results for $p/\epsilon$ and $s/n$ in a hadron
resonance gas. $p/\epsilon$ can be related to  $c_s$ at very high
temperatures, i.e., in an ideal gas. We notice that $p/\epsilon$  reaches
$0.14$ at the critical point. 

The production of sterile neutrinos occurs through collisions between active
neutrino gas and plasma of weakly interacting particles. The production in
cold, warm and hot dark matter scenarios~\cite{dm1} is maximum around $\sim
130~$MeV. Intuitively, one can think about a close relation between sterile
neutrinos and QCD phase transition. According to~\cite{kevork}, first order
phase transition significantly enhance sterile neutrino relic densities
relative to crossover one. 

\begin{figure}[thb]
\centerline{
\includegraphics[width=9.cm]{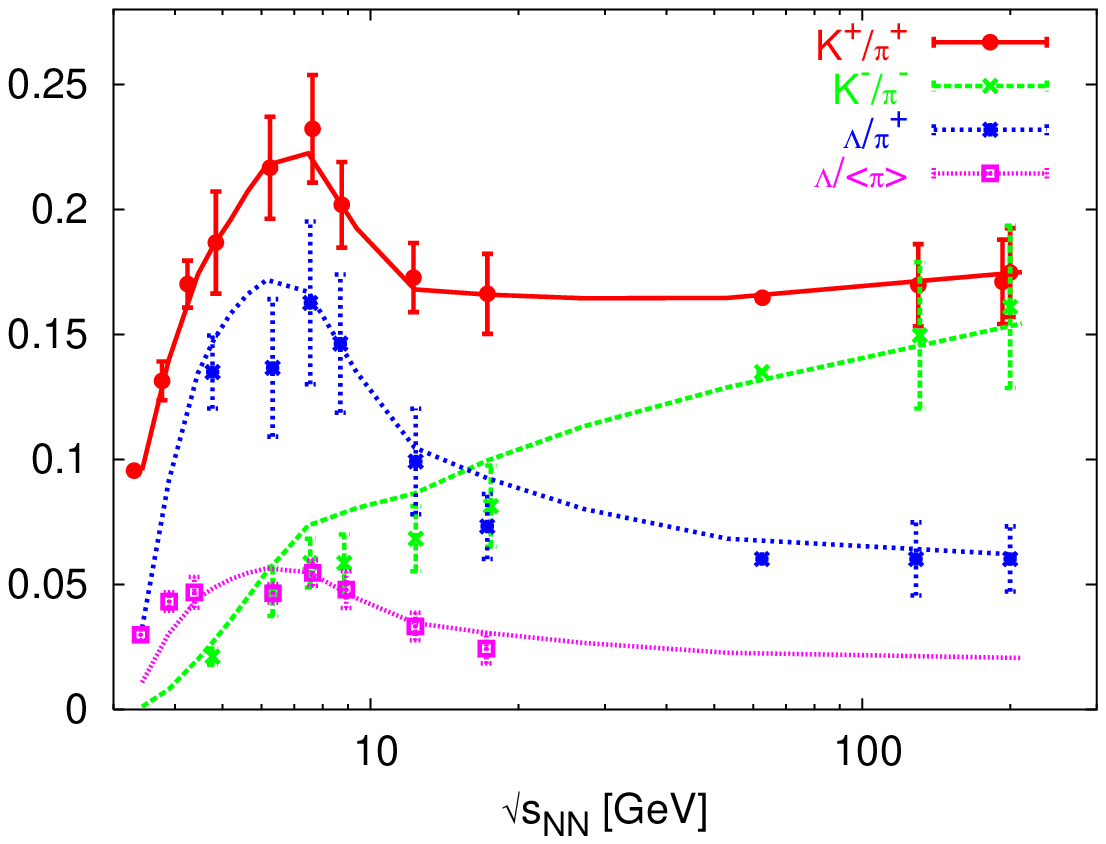}
\includegraphics[width=9.cm]{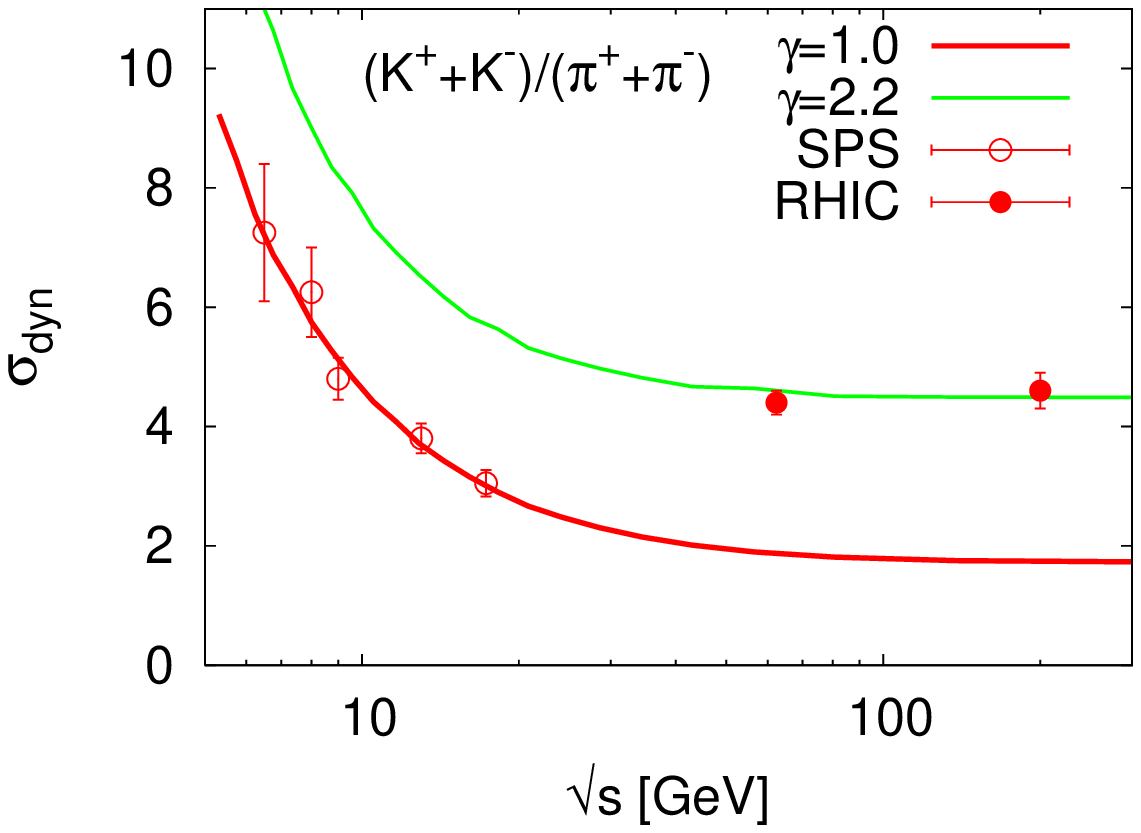}
}
\caption{Left panel: ratios of different particles as functions of
  $\sqrt{s}$~\cite{taw11}. The largest  $\sqrt{s}$ is related to 
  early universe, where the net baryon number is almost zero. Right panel:
  the dynamical fluctuations of $K/\pi$ particle ratios
  $\sqrt{s}$~\cite{taw22}. The points give the experimental data from
  various heavy ion collisions. The curves show the results from a hadron
  resonance gas~\cite{TawLat}.}  
 \label{Fig:eufig1}  
\end{figure}

QCD predicts another phase transition at almost same $T_c$ of
confinement-deconfinement phase transition; the restoration of chiral breaking
symmetry, i.e., anisotropic rotation of right and left handed quarks. The
breaking chiral symmetry results in the pions as the lightest Goldstone
bosons. Therefore, strong fluctuations on the pion fields known as DCC, {\it
  disordered chiral condensates}~\cite{dcc1}, are quite likely through the
transition. If chiral phase transition takes place through an
out-of-equilibrium process, it would provide a crucial tool to explain the
primordial magnetic fields~\cite{dcc2}. The origin of the primordial magnetic
fields shall be studied by the Square Kilometer Array (SKA)~\cite{dcc3}. The
particle production can be studied in hadron resonance gas (below $T_c$). Left
panel of Fig,~\ref{Fig:eufig1} shows ratios of various particles as a function
of $\sqrt{s}$~\cite{taw11}. 

The dynamical fluctuations associated with strong first order of phase
transition are likely very large. The continues second order or cross over
phase transition might wish out large part of dynamical fluctuations in the
final state and do not provide out-of-equilibrium.  
On the other hand, the dynamical fluctuations are conjectured to slow down
near the second order phase transition. This has been confirmed in classical
systems, solid state physics. In quantum field theory, the long-wavelength
(spinodal) modes will be quenched through the second order phase
transition~\cite{quench}. 

In right panel of Fig,~\ref{Fig:eufig1}, we notice that the dynamical
fluctuations smoothly increase with the center-of-mass energy $\sqrt{s}$. This
agrees with lattice simulations for $n_q$ fluctuations~\cite{karsch22} and
might support the conclusions that any anomalous phenomenon associated with
non-equilibrium phase change likely would be washed out, if the phase change
is smooth and does not cause out-of-equilibrium.

In second order phase transition, the correlation length $\zeta$ likely
diverges at the critical point. This divergence can be characterized by a
critical exponent $\nu$. $\nu$ has a universal value, $\sim2/3$. It does not
depend on the microscopic details of the system in transition. It only depends
on the universality class. 
\ba \label{zetaa1}
\zeta(T) &=& \zeta(T_c)\left(\frac{T_c-T}{T_c}\right)^{-\nu}
\ea 
The relaxation time $\tau$ also diverges at the critical point. $\tau$ gives
the time needed for a small perturbation to be in-equilibrium.
\ba \label{zetaa2}
\tau &=& \tau_0\left(\frac{T_c-T}{T_c}\right)^{-\mu}
\ea 
Assuming that the system cools down with a constant rate so that 
\ba \label{zetaa3}
T(t) &=& T_c \left(1-\frac{t}{\tau_Q}\right)
\ea 
where $\tau_Q$ is the quench time. 
From Eq.~(\ref{zetaa1}), (\ref{zetaa2}) and  (\ref{zetaa3}),
the correlation length and relaxation time can be related to $t/t_Q$
\ba
\zeta(t) &=& \zeta_0\left(\frac{t}{\tau_Q}\right)^{-\nu} \\
\tau(t) &=& \tau_0\left(\frac{t}{\tau_Q}\right)^{-\mu}
\ea
The fluctuations that survive the phase transition are depending on the freeze
out time. The fluctuations are characterizing these fluctuations. The freeze
out time can be obtained by solving the equation $\tau(\hat{t})=\hat{t}$ 
\ba
\hat{t} &=&-\left(\tau_0\tau_Q^{\mu}\right)^{1/(1+\mu)}
\ea

As shown above, the hydrodynamic evolution of viscous QGP fluid is slower than
the ideal QGP, i.e. $\tau(t)>t$ or any out-of-equilibrium process is slow and
therefore can survive the phase transition. That we do not observe dynamical
fluctuations would indicate that the phase transition is continuous
(cross over), Fig.~\ref{Fig:eufig1}.
 
\subsection{Chemical freeze out}
At chemical equilibrium, $s/T^3=7$ for three quark flavors and the freeze out
temperature $T_{fo}\approx 174~$MeV.  $s/T^3=7$ is a universal condition
describing all experimental data from heavy ion collisions at a wide range on
incident energies~\cite{TawRt}. Then the entropy in QCD epoch reads 
\ba
S = V \cdot T^3 \approx 2.05 \cdot 10^{58}
\ea
This number reflects that about ten orders of magnitude an increase in
produced particles as a reason of QCD phase transition are expected. The
multiplicities of produced particles in heavy ion collisions should manifest
this increase. Should we take into consideration that the Universe meanwhile
expands this number becomes larger.

\subsection{Non-ideal QGP (strongly correlated QGP)}
According to~\cite{chaudhuri}, the hydrodynamic evolution of QGP fluid with
dissipation due to shear viscosity has a slower rate than the ideal QGP
fluid. The transverse expansion in the viscous fluid is stronger and faster
than the one in an ideal fluid. The pion production is considerably enhanced in
viscous fluid, so that larger viscosity straightforwardly leads to more
pions. The particle production increases when the freeze out surface is
extended and the distribution function become non-equilibrium.  The freeze out
surface~\cite{pasi1} can be parameterized and fitted to the experimental
data. With extension we mean change. The conditions deriving the freeze
out as a function of the incident energies are discussed in~\cite{TawRt}.  

The dissipation has another crucial consequence. It might modify the elliptic
flow. The elliptic flow reduces with the viscosity. The reduction tends to
reach a saturated value at high energies (large transverse momentum). The
ideal fluid shows almost opposite behavior. The elliptic flow increases with
the transverse momentum.     

\section{Summary and Conclusion}

Lattice QCD calculations with dynamical quarks and physical masses show that
the phase transition at very small net baryon density, which is related to
early universe is likely cross over or very weak second
order. Phenomenological indications for such a continuous transition is quite
likely an ambiguous task. As discussed earlier, the order and critical
temperature of deconfinement-confinement phase transition(s) calculated in  
lattice QCD strongly depend on the number and masses of quark flavors. The
order varies from strong first order to very weak second order or cross
over. Consequently, the critical temperature takes values between $150~$MeV
and $200~$MeV. The cosmological consequences strongly depend on the QCD phase
transition(s).  

We used to assume that QGP can be treated as a free ideal gas. The recent RHIC
results suggested that QGP is a fluid rather than an ideal gas. Also recent
lattice QCD simulations support such a conclusion. Taking into consideration
that quark-gluon plasma is strongly correlated (sQGP) and has shear viscosity
would lead to the 
consequence that its hydrodynamic evolution is slower relative to QGP with
zero viscosity. Also the transverse expansion in such a fluid is much stronger
than the one in an ideal QGP fluid. Phenomenologically, the particle
production is considerably enhanced in sQGP. All these results would to some
extent modify our ideas about the time evolution in early universe, for
example.

\end{document}